\documentstyle[epsf]{mn}

\newif\ifAMStwofonts
\def\amin{\ifmmode ^{\prime}\else$^{\prime}$\fi}
\def\asec{\ifmmode ^{\prime\prime}\else$^{\prime\prime}$\fi}
\newcommand{\be}{\begin{equation}}
\newcommand{\ee}{\end{equation}}
\newcommand{\bd}{\begin{displaymath}}
\newcommand{\ed}{\end{displaymath}}
\newcommand{\bea}{\begin{eqnarray}}
\newcommand{\eea}{\end{eqnarray}}

\newcommand{\gapprox}{\;\rlap{\lower 2.5pt
             \hbox{$\sim$}}\raise 1.5pt\hbox{$>$}\;}
\newcommand{\lapprox}{\;\rlap{\lower 2.5pt
             \hbox{$\sim$}}\raise 1.5pt\hbox{$<$}\;}

\ifoldfss
  \ifCUPmtlplainloaded \else
    \NewTextAlphabet{textbfit} {cmbxti10} {}
    \NewTextAlphabet{textbfss} {cmssbx10} {}
    \NewMathAlphabet{mathbfit} {cmbxti10} {} % for math mode
    \NewMathAlphabet{mathbfss} {cmssbx10} {} %  "   "    "
  \fi
  \ifAMStwofonts
    \ifCUPmtlplainloaded \else
      \NewSymbolFont{upmath} {eurm10}
      \NewSymbolFont{AMSa} {msam10}
      \NewMathSymbol{\upi}     {0}{upmath}{19}
      \NewMathSymbol{\umu}     {0}{upmath}{16}
      \NewMathSymbol{\upartial}{0}{upmath}{40}
      \NewMathSymbol{\leqslant}{3}{AMSa}{36}
      \NewMathSymbol{\geqslant}{3}{AMSa}{3E}

    \fi
  \fi
\fi % End of OFSS

\ifnfssone
  \newmathalphabet{\mathit}
  \addtoversion{normal}{\mathit}{cmr}{m}{it}
  \addtoversion{bold}{\mathit}{cmr}{bx}{it}
  \newmathalphabet{\mathbfit} % math mode version of \textbfit{..}
  \addtoversion{normal}{\mathbfit}{cmr}{bx}{it}
  \addtoversion{bold}{\mathbfit}{cmr}{bx}{it}
  \newmathalphabet{\mathbfss} % math mode version of \textbfss{..}
  \addtoversion{normal}{\mathbfss}{cmss}{bx}{n}
  \addtoversion{bold}{\mathbfss}{cmss}{bx}{n}
  \ifAMStwofonts
    \ifCUPmtlplainloaded \else
      \UseAMStwoboldmath
      \makeatletter
      \new@mathgroup\upmath@group
      \define@mathgroup\mv@normal\upmath@group{eur}{m}{n}
      \define@mathgroup\mv@bold\upmath@group{eur}{b}{n}
      \edef\UPM{\hexnumber\upmath@group}
      \new@mathgroup\amsa@group
      \define@mathgroup\mv@normal\amsa@group{msa}{m}{n}
      \define@mathgroup\mv@bold\amsa@group{msa}{m}{n}
      \edef\AMSa{\hexnumber\amsa@group}
      \makeatother
      \mathchardef\upi="0\UPM19
      \mathchardef\umu="0\UPM16
      \mathchardef\upartial="0\UPM40
      \mathchardef\leqslant="3\AMSa36
      \mathchardef\geqslant="3\AMSa3E
    \fi
  \fi
\fi % End of NFSS release 1

\ifnfsstwo
  \DeclareMathAlphabet{\mathbfit}{OT1}{cmr}{bx}{it}
  \SetMathAlphabet\mathbfit{bold}{OT1}{cmr}{bx}{it}
  \DeclareMathAlphabet{\mathbfss}{OT1}{cmss}{bx}{n}
  \SetMathAlphabet\mathbfss{bold}{OT1}{cmss}{bx}{n}
  \ifAMStwofonts
    \ifCUPmtlplainloaded \else
      \DeclareSymbolFont{UPM}{U}{eur}{m}{n}
      \SetSymbolFont{UPM}{bold}{U}{eur}{b}{n}
      \DeclareSymbolFont{AMSa}{U}{msa}{m}{n}
      \DeclareMathSymbol{\upi}{0}{UPM}{"19}
      \DeclareMathSymbol{\umu}{0}{UPM}{"16}
      \DeclareMathSymbol{\upartial}{0}{UPM}{"40}
      \DeclareMathSymbol{\leqslant}{3}{AMSa}{"36}
      \DeclareMathSymbol{\geqslant}{3}{AMSa}{"3E}
    \fi
  \fi
\fi % End of NFSS release 2

\ifCUPmtlplainloaded \else
  \ifAMStwofonts \else % If no AMS fonts
    \def\upi{\pi}
    \def\umu{\mu}
    \def\upartial{\partial}
  \fi
\fi

\newcommand{\rmd}{\mathrm{d}} 
\newcommand{\bftem}{}

\title{The X-ray spectrum of a disk illuminated by ions}
\author[H.\ C.\ Spruit and F.\ Haardt]
       {H. C. Spruit$^1$ and F.\ Haardt$^2$ \\
        $1$Max-Planck-Institut f\"ur Astrophysik, \\
        Postfach 1523, D-85740 Garching, Germany\\
        $^2$Universita' dell'Insubria, Dipartimento di Scienze\\
       via Lucini 3, 22100 Como, Italy}

\date{Accepted xx
      Received ;
      in original form xx}

\pagerange{\pageref{firstpage}--\pageref{lastpage}}
\pubyear{2000}

\begin{document}

\maketitle

\label{firstpage}

\begin{abstract}
The X-ray spectrum from a cool disk embedded in an ion supported torus is computed. {\bftem
 The interaction of the hot ions with the disk increases the hard X-ray luminosity of the system}. A surface layer of the disk is heated by the protons from the torus. The Comptonized  spectrum produced by this layer has a shape that depends only weakly on the  incident energy flux and the distance from the accreting compact object. It consists of a `blue bump' of unComptonized soft photons and a flat high energy tail, 
reminiscent of the observed spectra. {\bftem
 The hard tail becomes flatter as the 
thermalization depth in the cool disk is increased.  Further} evidence for ion illumination are the  Li abundance in the secondaries of low mass X-ray binaries and the 450 keV lines sometimes seen in black-hole transient spectra. 
\end{abstract}


\section{Introduction}

The X-ray spectra $F(E)$ of AGN as well as galactic black hole candidates (BHC) in 
their hard states are characterized by a power law of index $\approx 1$ and a high 
energy cutoff $E_{\mathrm c}$ around 200 keV {\bftem
 (e.g. Mushotzky et al. 1993, Ulrich et al. 1997, Cappi et al. 1997, Zdziarski et al. 1996)}. Such spectra are well 
known to be 
describable by Comptonization in an electron scattering layer of optical depth 
$\tau\approx 0.5$ and temperature $T\sim E_{\mathrm c}/2$. One of the classsical 
problems 
in X-ray astronomy is to explain why $\tau$ and $T$ should have just these values, 
with little variation between sources. Theoretical arguments can be given that 
Comptonization is in fact the most important interaction between matter and radiation 
at temperatures of 10--100 keV (e.g. Shakura and Sunyaev 1973), for the inferred 
radiation energy densities near accreting black holes, but this does not tell us what 
the the thickness and temperature of the interaction region are. 

An optically thick accretion disk would produce spectra peaking at 1 keV and 10--100 
eV for BHC and AGN, repectively. The conditions in BHC and AGN allow (at accretion 
rates well below Eddington) for a second form of accretion, an ion supported 
advection torus (Shapiro et al. 1976, Liang 1979, Ichimaru 19xx, Rees et al. 1982, Abaramowicz et al. 1988, Narayan and Yi 1994, Fabian and Rees 1995, Narayan et al. 1995, 1996). The ions in this flow are 
near their virial temperature, the electrons much cooler because of their weak 
interaction with the ions and their strong interaction with the radiation field. Such 
flows could produce, in principle, the kind of spectrum observed (Narayan et al. 
1995, 1996), but {\bftem
 more physics must be invoked to restrict} the optical depth and  temperature of the flow to the observed narrow ranges (cf. Haardt 1997, Maraschi and Haardt 1997).

\subsection{Evaporating disks inside tori}
Various geometries for the accretion flow near a black hole have been developed, for 
a review see Collin (1997). One of the possibilities is an ion supported advection 
torus coexisting with an optically thick accretion disk embedded in it (see fig. 1 in 
Collin). This possibility is attractive because the spectra of BHC in their high 
states show evidence of the simultaneous presence of an optically thick, thermal, 
accretion disk and a hotter component which produces a power law tail at higher 
photon energies (e.g. Mitsuda et al. 1984, Tanaka 1997). 

Theoretically, one would expect exchange of both mass an energy to take place between 
the disk and the advection torus. Heating of the disk surface by the hot ion 
supported flow above would lead to an `evaporation' of the disk surface, feeding mass 
into the torus. Such evaporation has been studied in detail for the case of disks in 
Cataclysmic Variable systems by Meyer and Meyer-Hofmeister (1994). In the inner 
regions of the disk the mass available in the disk is smallest, and the energy budget 
potentially available for evaporation largest. If the mass flow from disk into torus 
increases with the energy dissipation rate, and if a steady state develops, one could 
therefore envisage a structure consisting of three regions: an outer one in which 
only a geometrically thin optically thick disk is present, inside this a composite 
region with an evaporating disk inside a hot ion supported advection torus, and 
inside this a region in which only an ion supported flow exists because all disk mass 
has evaporated (Meyer and Meyer-Hofmeister 1994, Meyer-Hofmeister and Meyer 1999). 

Depending on details of the processes of mass and energy exchange between disk and 
torus, the boundaries between these regions may vary. It is not necessary that the 
structure is steady. The model has, in principle, sufficient ingredients to allow for 
variability and may perhaps be developed further in the context of the various forms 
of variability seen in BHC. 

\subsection{Energy exchange between disk and torus}
Energy exchange between disk and torus may be mediated by particles or by radiation. {\bftem
 In its simplest form of interaction, the torus provides hard photons that are reprocessed by the disk, a model used extensively  for interpretations of X-ray spectra (e.g. Wandel and Liang 1991,  Reynolds et al. 1994, Petrucci and Henri 1997, Gilfanov et al. 1999 and references therein). 

A more internally consistent model for this interaction is that of Haardt and  Maraschi (1991). To simplify the discussion, assume that the accretion takes place 
predominantly through an ion torus or a hot `corona' above the disk} (this assumption 
can easily be relaxed). The 
radiation produced by the torus illuminates the disk below, which thermalizes it into 
an approximate blackbody spectrum. These (soft) photons are Comptonized in the hot 
torus. In this model, approximately half the energy comes out as soft radiation and 
half as Comptonized photons. It correctly predicts the slope of the spectrum, {\bftem
 and fixes a relation between the temperature and  optical depth of the Comptonizing layer.  {\bftem
 An assumption about the temperate of of Comptonizing region is still needed to 
get the right cutoff energy $E_{\mathrm c}$, though pair ceation limits the possible temperatures}}. 
For further developments of this model see  Haardt (1997).

A second channel of energetic interaction is the hot protons in the torus with 
temperature near the virial temperature, $T_{\mathrm p}\approx T_{\mathrm v}\approx 
160 
r_{\mathrm g}/r$ MeV. At the distance dominating the energy release, $r\approx 
7r_{\mathrm 
g}$, the protons thus have a temperature around 20 MeV. At this energy, they have a 
significant penetration depth into the cool disk. They are slowed down mainly by 
Coulomb interactions with the ensemble of electrons inside their Debye sphere. The 
`stopping depth', expressed in terms of the corresponding Thomson optical depth, is 
\be
\tau_{\mathrm s}\approx {m_{\mathrm p}\over 3 m_{\mathrm 
e}\ln\Lambda}{\beta^4\over 
\psi-x\psi^\prime}, \label{lsr}
\ee
(e.g. Ryter et al., 1970) where $\beta=v_{\mathrm z}/c\approx (kT_{\mathrm 
p}/m_{\mathrm 
p}c^2)^{1/2}$ is the vertical component of the proton velocity, $\theta=kT/m_{\mathrm 
e}c^2$ is a measure of the temperature of the heated layer, $\ln\Lambda\approx 20$ is 
a Coulomb logarithm, $x^2=\beta^2/(2\theta)$, and $\psi$ the error function
\be \psi={2\over\sqrt\upi}\int_0^xe^{-x^2}\rmd x. \label{ls}\ee
This formula holds for nonrelativistic temperatures; the relativistic
generalization has been given by Stepney (1983) and Stepney and Guilbert 
(1983).    At low temperature $kT<m_{\mathrm e}/m_{\mathrm 
p}\,kT_{\mathrm p}$, $x$ is 
small and the factor involving the error function can be expanded. This yields
\bea 
\tau_{\mathrm s}\approx &{m_{\mathrm p}\over m_{\mathrm e}\ln\Lambda} 
\beta\theta^{3/2}, \cr 
\approx &\left(kT_{\mathrm p}\over 50 {\mathrm MeV}\right)^{1/2}
\left(kT\over 50 {\mathrm keV}\right)^{3/2}.\label{lsa}
\eea

\section{Comptonization in a layer heated by protons}

\subsection{Estimating the depth of the Comptonizing layer}
Heating by protons yields a Comptonizing layer of thickness equal to the stopping 
depth $\tau_{\mathrm s}$. This depth is a function of the temperature in the 
layer, by (\ref{lsr}). The temperature on the other hand is determined by 
the heating and cooling processes, so that a consistent calculation of heating and 
cooling will yield both the temperature and the optical depth of the layer. 
With a simple estimate, we can now show that this will yield $\tau_{\mathrm s}$ and 
$T$ 
in roughly the right range. 

The cooling process in the layer is the inverse Compton process, i.e. the energy loss 
electrons experience as they scatter the soft photons from the cool disk below. We 
assume that these soft photons are all (or mostly) produced by thermalization of 
Comptonized photons from the heated layer, as in the model of Haardt and Maraschi 
(1991, hereafter HM). Since approximately half the Comptonized photons escape and the 
other half illuminates the thermalizing layer, the energy flux in the soft photons at 
the base of the layer must be about the same as that in the escaping Comptonized 
photons. Such a balance is possible only if the Comptonization is sufficiently 
strong. In terms of the Compton $y$-parameter $y\approx 4\theta\tau_{\mathrm s}$, it 
requires that $y\approx 1$. If the temperature is too low, the energy 
transfer
from the electrons to the soft photons is too low and the layer heats up until 
$y\approx 1$, and vice versa. Since the $y$-parameter also determines the slope of 
the X-ray spectrum, the model yields a fixed spectral slope, which is in the range of 
the observed values. 
%This is the reason for the success of the HM model. 
In HM the 
depth of the layer is a free parameter; in the present model, it is fixed by 
requiring the stopping depth $\tau_{\mathrm s}$ to be consistent with the resulting 
temperature $T$. A simple estimate is obtained by setting $y=1$, or 
$\theta=1/(4\tau_{\mathrm s})$, and inserting into (\ref{lsr}). This yields, assuming 
again that the protons are near their virial temperature:
\be 
\tau_{\mathrm s}^{5/2}={m_{\mathrm p}\over 8\sqrt{6}\ln\Lambda m_{\mathrm 
e}}\left({r_{\mathrm 
g}\over r}\right)^{1/2},
\ee
or
\be \tau_{\mathrm s}\approx 1.3 \left({7r_{\mathrm g}\over r}\right)^{1/5}, \ee
and
\be 
kT\approx m_{\mathrm e}c^2/(4\tau_{\mathrm s})\approx 60 \left({r\over 
7r_{\mathrm 
g}}\right)^{1/5} \quad {\mathrm keV}.
\ee

We conclude that proton illumination yields optical depths and temperatures in the 
right range, with only a weak dependence on the assumed distance from the black hole. 
Obviously, the estimate is rather crude, and more detailed calculations of the energy 
transfer from the protons to the electrons, as well as the Comptonization process are 
needed to test the model. In the following we make a first step in this direction, by 
means of a radiative transfer calculation.

\subsection{Model problem}
The aim of the calculation reported below is to compute the temperature as a 
function of depth, together with the emergent photon spectrum from a layer heated by 
protons who deposit their energy according to (\ref{lsr}). A preliminary account of the calculation has been given in Spruit (1997).

We approximate the heating rate to be distributed uniformly over a layer with depth 
$\tau_{\mathrm s}$, where $\tau_{\mathrm s}$ is computed from the average 
temperature in this 
layer using eq. (\ref{lsr}). I reality, the heating is somewhat non-uniform because 
of the dependence of the energy loss rate of the proton on both the proton velocity 
and the temperature. The approximation of a uniform energy input in the layer 
is deemed sufficient for the present exploratory purpose. 

In this layer, we solve the radiative transfer equation iteratively together with the 
temperature $T(\tau)$, and the layer stopping depth $\tau_{\mathrm s}$ such 
that 
the heating is in balance with cooling by Comptonization of the soft photons. 

The assumptions and simplifications that go into the model are as follows. In the 
heated layer, the only photon process considered is electron scattering 
(Comptonization). Below this, we assume that electron scattering continues to be the 
dominant process down to some depth $\tau_{\mathrm b}$. At $\tau_{\mathrm b}$, the 
downward 
photons are assumed to be absorbed and their energy reradiated upward as a black body 
spectrum. Thus, the gradual thermalization with depth by free-free processes is 
simplified by a step at depth $\tau_{\mathrm b}$. The value of $\tau_{\mathrm s}$ is 
determined from the proton velocity $\beta=v/c$ and the mean temperature in 
the heated layer. Obvious improvements are possible on these simplifications, by 
explicitly taking into account photon production/destruction process, an by a more 
accurate treatment of the energy loss of the ions as they penetrate into the disk.

The radiative transport part of the problem is simplified by reducing the angular 
dependences to a one-stream model: only vertically upward and downward moving 
photons 
are considered, and the electron distribution is similarly reduced from a 3-D to a 
one-dimensional Maxwellian distribution. For the scattering cross section and the 
electron Maxwellian the relativistic expressions are used. The one-stream 
simplification is made for programming convenience only: leaving out the full angular 
dependences leads to very simple expressions. Discretization by a reasonable number 
of angles would still yield a very modest problem in terms of computing time, and is 
an obvious next step to improve the calculations.

\subsection{Numerical method}
The transport equation is of the form (e.g. Rybicki and Lightman, 1976):
\bea 
{\rmd n\over\rmd z} = &\int \rmd^3 {\mathbf p}\int \rmd^2 \mathbf{\Omega} 
{\rmd \sigma \over \rmd \mathbf{\Omega}} [f_{\mathrm e}({\mathbf p}^\prime )n 
(\mathbf{\omega}^\prime)\times \cr
&(1+n(\mathbf{\omega})) - f_{\mathrm e}({\mathbf p}) n (\mathbf{\omega}) 
(1+n(\mathbf{\omega}^\prime))],
\eea
where $n({\mathbf\omega})$ is the photon occupation number, $f_{\mathrm e}$ the 
electron 
momentum distribution, ${\mathbf p}$ (respectively ${\mathbf p}^\prime$) the electron 
momentum, $\mathbf{\omega}$ ($\mathbf{\omega}^\prime$) the photon momentum 
vectors before 
(after) scattering, and $\mathbf{\Omega}$ the scattering angle. The dependences of 
${\mathbf 
p}^\prime$ and $\mathbf{\omega}^\prime$ on (${\mathbf 
p},\mathbf{\omega},\mathbf{\Omega}$) follow 
from the collision kinematics.

This equation is discretized in $N_\omega$ logarithmically spaced photon-energy bins. 
As photon energy scale we use $\omega=h\nu/m_{\mathrm e}c^2$. As depth scale we 
use the 
Thomson optical depth $\tau_{\mathrm T}$. The number of depth points is fixed, but the 
grid is stretched such that at each iteration the stopping depth $\tau_{\mathrm s}$ is 
located at the same grid point. Half the grid points are used for the heated layer 
($0<\tau<\tau_{\mathrm s}$), the other half for the scattering layer below.

If $n_i^+$ and $n_i^-$ are the occupation numbers of the upward and downward moving 
photons in bin $i$, the result is the set of $2N_\omega$ equations
\be 
\mp{\rmd 
\over\rmd\tau}n_i^\pm=n_i^\pm\sum_j[(B_{ji}-B_{ij})n_j^\mp-B_{ij}]+\sum_j 
B_{ji}n_j^\mp, \label{treq}
\ee
where $B_{ij}(T)$ is the scattering cross section integrated over the appropriate 
frequency and electron momenta, for scattering from bin $i$ into bin $j$, in units of 
the Thomson cross section $\sigma_{\mathrm T}$.
At the top boundary there are only outgoing photons: 
\be n_i^-=0  \qquad(\tau=0), \ee
while at the lower boundary the downward photons are thermalized into a blackbody 
spectrum of upward photons:
\be n_i^+=n_{\mathrm BB}(\omega_i)=1/[1-\exp(\omega/\theta_{\mathrm b})]\qquad 
(\tau=\tau_{\mathrm b}).
\ee
Here $n_{\mathrm BB}$ is the black-body occupation number at temperature 
$\theta_{\mathrm 
b}$. This temperature follows from the condition that the net energy flux at depth 
$\tau_{\mathrm b}$ vanishes:
\be \int \omega^3[n^-(\tau_{\mathrm b})-n_{\mathrm BB}(\omega,\theta_{\mathrm 
b})]=0.\ee
This is a result of our assumption that the internal energy production rate in the 
cool disk can be neglected in comparison with the incident proton energy flux 
(generalizations are easily made by adding a term corresponding to the internal 
disspation in the disk).

The condition of energy balance between the assumed heating rate $h(\tau)$ and the 
Compton cooling by the soft photons is 
\be {\rmd F\over \rmd z}=h, \ee
where the energy flux $F$ is given by
\be F=\sum_i\omega^3(n_i^+-n_i^-). \ee

The equations (\ref{treq}) are discretized in optical depth by centered first order 
differences. The resulting set of nonlinear algebraic equations is solved by an 
iterative process. It turned out that full simultaneous linearization of the transfer 
equation, the boundary conditions and the energy equation had very poor convergence 
properties. Instead, an iteration was done in which only the transfer equation and 
the upper boundary condition were linearized, while the lower  boundary condition and the energy 
equation were dealt with by a modified succesive-substitution process after each 
iteration of the transfer equation. Convergence, however, was still problematic for 
large optical depths and for cases where the assumed photon energy range extended too 
far beyond the cutoff energy. For the cases reported here, where the optical depth is 
not too large, on the order of 30--100 iterations were required for an accuracy of 
$10^{-3}$ in luminosity.

\section{Results}

\begin{figure}
\mbox{}\hfil\mbox{\hspace{-1cm}\epsfysize 7.2cm\epsfbox{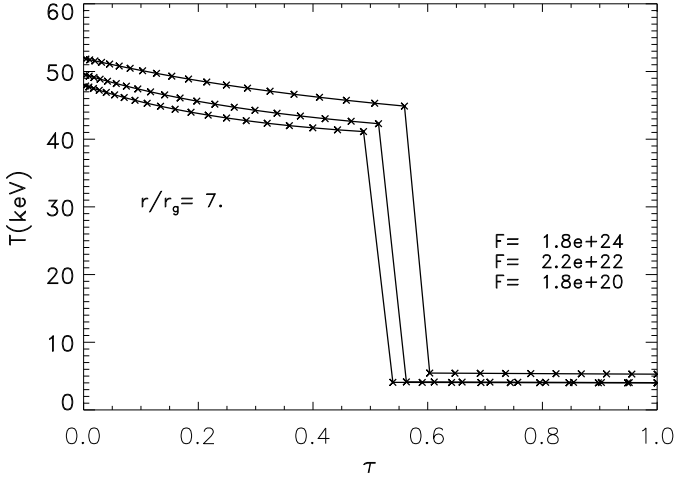}}\hfil\mbox{}
      \caption{Temperature as a function of optical depth in an electron scattering 
layer heated by virialized protons incident with energy flux $F$ (erg/cm$^2$/s), for 
$r/r_{\mathrm g}=7$. Lower boundary (thermalizing layer) is at $\tau=1$. Crosses mark 
grid points}
      \label{et}
\end{figure}

\begin{figure}
\mbox{}\hfil\mbox{\epsfysize 6.2cm\epsfbox{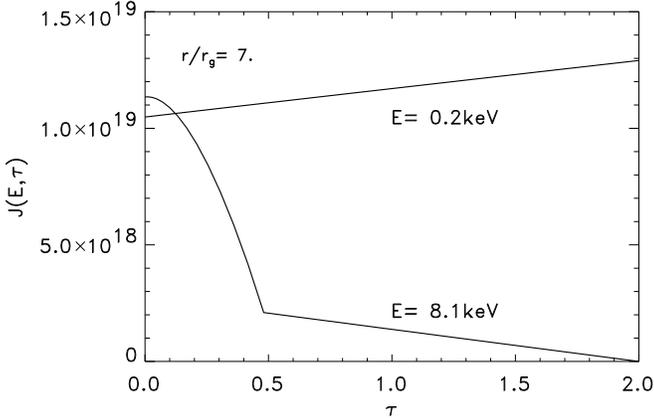}}\hfil\mbox{}
      \caption{Variation with depth of the mean intensity $J(E,\tau)$ of soft  
(0.2\,keV) and hard (8\,keV) photons. Energy flux $F=1.8\,10^{20}$, 
$r/r_g=7$.}
      \label{sz}
\end{figure}

\begin{figure}
\mbox{}\hfil\mbox{\hspace{-0.7cm}\epsfysize 6.5cm\epsfbox{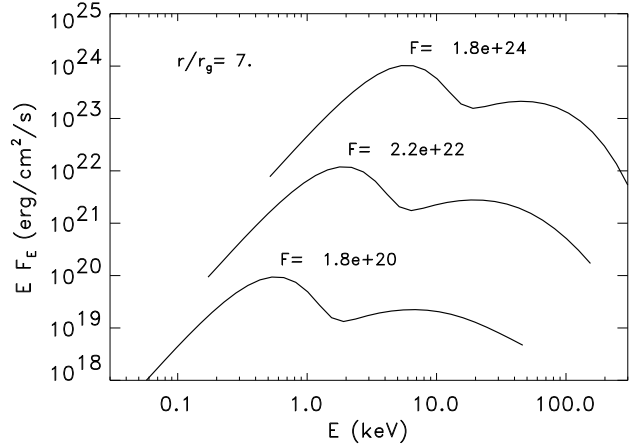}}\hfil\mbox{}
      \caption{Emergent spectrum of a layer heated by protons at $r=7r_{\mathrm g}$ for 
three values of the energy flux}
      \label{sf}
\end{figure}
\begin{figure}
\mbox{}\hfil\mbox{\hspace{-0.7cm}\epsfysize 6.7cm\epsfbox{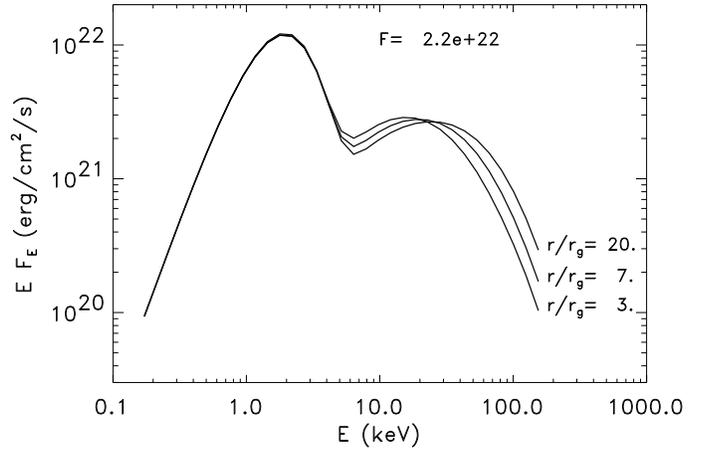}}\hfil\mbox{}
      \caption{Dependence of the spectrum on distance from the compact object, for a 
fixed energy flux per cm$^2$ of disk surface}
      \label{sr}
\end{figure}
\begin{figure}
\mbox{}\hfil\mbox{\hspace{-1cm}\epsfysize 7.2cm\epsfbox{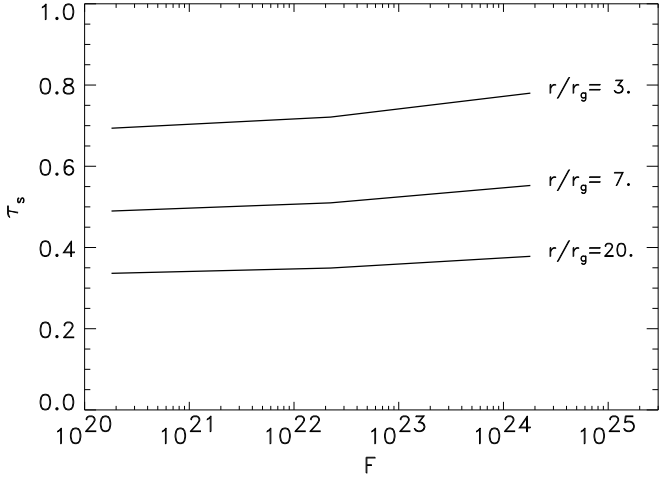}}\hfil\mbox{}
      \caption{Thomson depth of the ion-heated layer as a function of energy flux and 
distance from the compact object}
      \label{pen}
\end{figure}

The parameters of the problem are the total energy flux $F$ (per unit surface area of 
the disk), the velocity of the incident protons, and the optical depth $\tau_{\mathrm 
b}$ 
of the thermalizing lower boundary. Assuming the protons to be thermal and 
virialized, their mean vertical velocity component $\beta_z$ as a function of the 
distance $r$ from the compact object is
\be \beta_z=\left({r_{\mathrm g}\over 6r}\right)^{1/2}\ee
The temperature as a function of depth is shown in figure \ref{et} for a few 
values of $F$ for $r/r_{\mathrm g}=7$. These fluxes correspond to effective 
temperatures 
$\theta_{\mathrm eff}=(F/\sigma)^{1/4}/(m_{\mathrm e}c^2)$ of $3\,10^{-4}$, 
$10^{-3}$ and 
$3\,10^{-3}$, respectively. The emergent spectrum for these energy fluxes is shown 
in figure \ref{sr}, for $r/r_{\mathrm g}=7$. The dependence of the spectrum on 
$r/r_{\mathrm 
g}$ at a fixed $F$ is shown in figure \ref{sf}. The penetration depths are shown in 
figure \ref{pen}. All of these cases were computed for $\tau_{\mathrm b}=1$.

The jump in temperature at $\tau=\tau_{\mathrm s}$ is a consequence of the fact that 
the 
heating rate jumps at $\tau_{\mathrm s}$. The photon field is continuous at 
$\tau_{\mathrm 
s}$ as shown in figure \ref{sz}, as expected because of their diffusion through the 
scattering layer. In order to balance the energy input, the electrons have to loose 
enough energy by scattering soft photons. This requires them to be hot where the 
energy input rate is large, and causes the temperature to follow the heating rate.
The jump will be smoothed if a more realistic model for the proton penetration is used. 
If the protons are not monoenergetic and unidirectional but taken from a thermal 
distribution, for example, the transition in heating rate will be smoother.

\begin{figure}
\mbox{}\hfil\mbox{\hspace{-0.5cm}\epsfysize 6.7cm\epsfbox{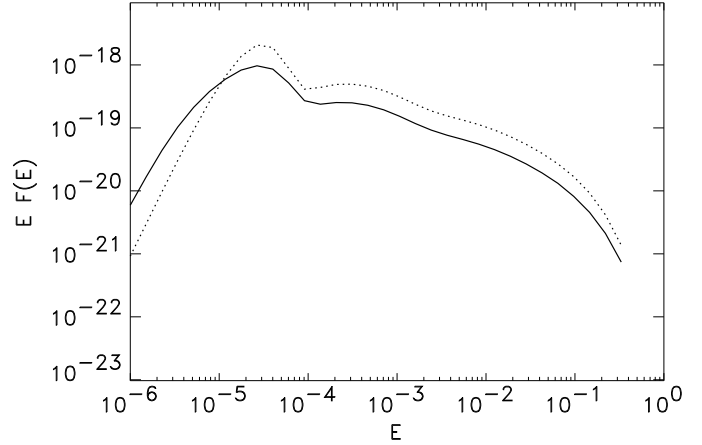}}\hfil\mbox{}
      \caption{Spectrum integrated over the disk (solid), compared with local 
spectrum at $r=7r_{\mathrm g}$ (broken).}
      \label{conv}
\end{figure}

The similarity of the spectra, apart from shifts in amplitude and photon energy, is 
remarkable. The temperature of the heated layer increases only very weakly with 
increasing energy flux. The cutoff energy increases somewhat with distance from the 
hole, but again this is dependence is rather weak. On account of the modest optical depth 
of the layer, the spectrum shows a prominent contribution from unscattered soft 
photons from the reprocessing depth $\tau_{\mathrm b}$. This peak is smeared out 
somewhat 
when the spectra are convolved over distance from the hole, at a given accretion 
rate. This is shown in figure \ref{conv}.

\begin{figure}
\mbox{}\hfil\mbox{\hspace{-0.5cm}\epsfysize 6.7cm\epsfbox{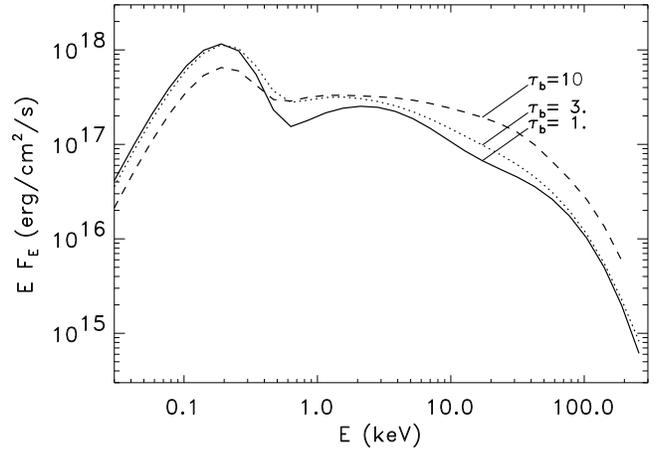}}\hfil\mbox{}
      \caption{Spectrum as a function of the depth of the thermalizing lower boundary 
$\tau_{\mathrm b}$, for an energy flux of $2.2\,10^{18}$ at $r=7r_{\mathrm g}$}
      \label{st}
\end{figure}

{\bftem
 The final parameter of the problem is the depth of the thermalizing lower boundary, 
$\tau_{\mathrm b}$. As this depth is increased, the thickness of the passive scattering 
layer between the thermalizing depth and the heated surface layer increases. This has two effects on the  spectrum. First, the strength of the `blue bump' (consisting of unComptonized thermal  photons from the lower boundary) decreases. Secondly the spectrum becomes flatter at the high energy end. This is shown in figure \ref{st}. With increasing depth of the thermalizing boundary, the spectrum becomes both smoother and flatter. 

This can be understood as a combination of two factors. First, the low energy photons 
produced at $z_{\rm b}$ must travel through a scattering layer before reaching the base 
of the Comptonizing layer, and for this a gradient in photon density is needed. On arrival 
at the base of the Comptonizing layer, the soft spectrum is therefore not a black body 
any more, but a `diluted black body'. For a given soft flux, there are fewer photons but of 
higher energy than if the thermalization were to take place directly at the base of the 
Comptonizing layer. Secondly, fewer Comptonized photons reach the thermalization 
layer, since the scattering layer in between effectively reflects them to some degree. 
The combination of these effects causes the output spectrum to be harder than if the Comptonizing layer makes direct contact with the thermalizing surface.

How much of the hardening effect remains if other radiation processes than pure scattering are taken into account requires more detailed calculations taking into account free-free and bound-free transitions. The thermalization  of the downward traveling Comptonized photons into soft photons is done by these processes, and represented only crudely by the thermalization at a fixed depth assumed here.}

\section{Discussion} 
With an admittedly somewhat simplified radiative transfer model I have shown that 
heating of a cool disk by protons from an ion supported advection torus produces 
X-ray spectra that are very reminiscent of the hard spectra of accreting black 
holes. Like the Maraschi and Haardt model (and for the same reason), it yields 
approximately the right spectral slope, but in addition, it also predicts 
approximately the right optical depth of the Comptonizing layer and the cutoff energy 
of the spectrum. 

The temperature of the heated layer is insensitive to the energy flux, and stays 
around 40--60 keV. Instead of getting hotter at high energy flux, the incident energy 
is spent in upscattering a larger number of soft photons. As in the Maraschi and 
Haardt model, the reason for this lies in the energy balance condition. In order for 
the incident energy to be radiated as Comptonized flux, the Compton $y$-parameter has 
to be of of the order unity, and the temperature of the layer adjusts accordingly. 
What is new in the results presented here is that the optical depth of the 
Comptonizing layer also comes out naturally in the right range, due to the physics of 
Coulomb interaction between the virialized protons with the electrons in the cool 
disk. This process also is fairly insensitive to the proton temperature (within the 
relevant range), so that the emergent spectrum is only a weak function of distance 
from the accreting object (figure \ref{sr}). 

The agreement of the result with one of the most puzzling features of the hard X-ray 
spectra of accreting compact objects, viz. the uniformity of the spectral shape, 
makes it likely that ion illumination plays a major role in the physics of these 
objects.

\subsection{Connection between disk and ADAF}
{\bftem
 The ion torus (ADAF) is underluminous for its accretion rate, because the ions do not have time to transfer their energy to the radiating electrons before being swallowed by the hole. The presence of a cool disk illuminated by the ions of the torus would increase the luminosity of the system. Since the illumination process produces a hard spectrum, like the torus itself, the presence of the cool disk can still be compatible with the observed hard spectra, but the accreting system would not be as underluminous as an ADAF without a cool disk. If any interaction at all takes place, the ADAF can not be underluminous by several orders of magnitude, as in some proposed ADAF applications. 

The ion energy lost by illumination acts as a cooling agent on the hot ion torus,  and limits the conditions under which it can exist. The hot ion supported flow has to be fed by a cool disk in some way or other, so that there must
be at least a small region where the two coexist and the illumination process takes place.  In the so-called intermediate state in black hole accreters (e.g. Rutledge etal), a soft and hard component coexist in the X-ray spectrum. If ion tori actually exist in these systems, this is observational evidence that disk and torus can coexist with both contributing significantly to the luminosity. }

\subsection{Lithium}
An independent observational indication for ion illumination is the observation of 
high Li abundances in the secondary stars of X-ray binaries (Mart\'{\i}n et al. 1992, 
1994a, 1995). The energy of virialized protons hitting the cool disk at $r=7r_{\mathrm 
g}$ (where the gravitational energy release peaks for accretion onto a hole) is 
around 50 MeV, just in the range where Li production by spallation of CNO elements 
becomes efficient. Since the observed secondaries are of spectral types known to 
destroy Li on a rather short time scale, a significant continual source of Li is 
needed. As shown by Mart\'{\i}n et al. (1994) the energetics of the accretion process 
is enough to explain the observed amount of Li on the secondary, if a fraction 
$10^{-3}$ of the Li produced in the disk finds its way to the secondary (in the form 
of a disk wind, for example). Another consequence of ion illumination would be Li and 
Be production by He nuclei from the virialized flow reacting with He nuclei in the 
disk. These reactions peak around 50 Mev/nucleon, and are accompanied with emission 
of $\gamma-$lines at 431 and 478 keV. It is possible that the $\gamma-$lines 
observed 
sometimes around this energy (Gilfanov et al. 1991, Sunyaev et al. 1992) are another 
signature of ion illumination (Mart\'{\i}n et al. 1994a,b).

\section*{Acknowledgments}
This work was done in the context of Human Capital and Mobility network `Accretion 
onto compact objects', CHRX-CT93-0329.

\bsp

\label{lastpage}

\end{document}